\documentclass[eqsecnum,preprint,12pt,superscriptaddress,aps]{revtex4}
\usepackage{epsfig}
\begin{document}

\title{A Fast Way to Compute  Functional Determinants of Radially Symmetric Partial Differential Operators in General Dimensions}
\author{Jin Hur}
\email{hurjin@kias.re.kr}
\affiliation{School of Computational Sciences, Korea Institute for Advanced Study, Seoul 130-012, Korea}
\author{Hyunsoo Min}
\email{hsmin@dirac.uos.ac.kr}
\affiliation{Department of Physics, University of Seoul, Seoul 130-743, Korea}
\affiliation{School of Physics, Korea Institute for Advanced Study, Seoul 130-012, Korea }

\vspace{3cm}
\begin{abstract}
Recently the partial wave cutoff method was developed as a new calculational scheme for a functional determinant of  quantum field theory in radial backgrounds. For the contribution given by an infinite sum of large partial waves, we derive explicitly radial WKB series in the angular momentum cutoff for $d=2,3,4$ and 5 ($d$ is the spacetime dimension), which has uniform validity irrespectively of any specific values assumed for other parameters. Utilizing this series, precision evaluation of the renormalized functional determinant is possible with a relatively small number of low partial wave contributions determined separately.  We illustrate the power of this scheme in numerically exact evaluation of the prefactor (expressed as a functional determinant) in the case of the false vacuum decay of 4D scalar field theory.
\end{abstract}
\maketitle

\section{Introduction}
Functional determinants of (ordinary or partial) differential operators arise in many areas of physics: for instance, in connection with the one-loop effective action in quantum field-theoretic studies and in the semiclassical approximation to quantum mechanical tunneling amplitudes. However, explicit evaluation of these quantities, especially with partial differential operators involving  nontrivial background fields, is usually a very difficult problem. Explicit  analytic results are known only in some simple cases, such as the one-loop effects in constant electromagnetic fields in QED \cite{heisenberg,duff,dunne-review} and in a covariantly constant field strength in non-Abelian gauge theories \cite{leutwyler,yildiz}. Therefore various methods for approximate calculation were considered,  the large mass expansion \cite{kwon} and the derivative expansion \cite{gargett,salcedo} being good examples of them. But the validity of these approximate methods crucially depends on the range of various parameters entering the problem.

Recently there has been a significant progress in this problem, at least when the differential operators are separable. Especially, for background fields having radial symmetry, a method using the partial wave analysis has been developed in the form of the partial wave cutoff method \cite{idet,radial1}. This method was first used in the computation of QCD instanton determinant \cite{idet} for an arbitrary value of quark mass. (The same quantity with massless quarks was calculated in a classic paper by 'tHooft \cite{thooft} long time ago). It was then applied to the evaluation of the one loop effective action for more general classes of radial background fields \cite{radial1}. The prefactor in the false vacuum decay rate, which requires an evaluation of the functional determinant also, is calculated by the same method \cite{dunnemin,baacke}. 

Of crucial importance in the above-mentioned calculational scheme is to find a simple way to extract a finite renormalized quantity from the infinite sum of partial wave contributions. In \cite{idet} this was achieved by introducing a cutoff in the partial wave sum and then finding a uniform radial WKB expansion for the sum of partial waves beyond the cutoff value (which is combined with the conventional renormalization counterterms). Combining the leading terms of this WKB expansion with the contribution from partial waves below the cutoff value, it is possible to secure a finite renormalized value in the limit of large cutoff value. [In \cite{dunnekirsten}, similar results were obtained using the zeta function technique]. More recently it is observed that the inclusion of higher order terms in the uniform radial WKB expansion greatly improves the rate of convergence for the infinite sum of partial wave contributions \cite{radial2}.  The computational labour needed for the functional determinant calculation is thus much reduced. Efficiency of this scheme will become especially conspicuous for functional determinants of higher-dimensional differential operators, thus making it an effective tool also for the studies of higher-dimensional quantum field theories.

In the present paper we will give a simplified derivation of the uniform WKB expansion and provide explicitly several leading terms of this expansion (needed for fast precision evaluation of  functional determinants) in general contexts. 
It is our hope that these explicit results  find useful applications in various related  problems. This paper is organized as follows. In Section \ref{setupsection} the partial wave cutoff method is explained briefly  and the desired form of the asymptotic WKB series is presented. In Section \ref{wkbsection}, after introducing the proper-time representation of the radial functional determinant, we derive the large $l$ expansion of the proper-time Green function. In Section \ref{radialsection} the infinite sum of  contributions from  high angular momentum is explicitly evaluated using the radial WKB series and the Euler-Maclaurin summation method and then the uniform WKB expansions (as descending series in the angular momentum cutoff $L$) are presented in various dimensions i.e., for $d=2,3,4,5$. 
Our formulas for the renormalized functional determinants have definitely faster convergence property compared, say, to those given in \cite{dunnekirsten}. In the next section we consider  a direct application of these results, finding a more accurate value for the false vacuum decay rate in the context of 4D scalar field theory. In Section \ref{gaugesection} we consider the functional determinant in gauge theories, where the radial potential has a linear dependence in angular quantum number $l$. But this does not change the general structure, and in this case also the appropriate coefficient functions in the uniform WKB expansion can be found  in explicit forms.

\section{Setting up the Problem} \label{setupsection}
In order to set the problem precisely, let us start by considering a pair of partial differential operators
\begin{eqnarray}
{\cal M} = -\partial^2 +V(r), \qquad\qquad {\cal M}^{\rm free} = -\partial^2,  \label{op}
\end{eqnarray}
where $\partial^2=\partial_\mu \partial_\mu$ is the Laplace operator in $d$ dimension and $V(r)$ is a radial potential vanishing sufficiently fast at infinity. In the one-dimensional case (i.e., with ${\cal M}=-\frac{d^2}{dr^2}+V(r)$) with the Dirichlet boundary condition on the interval $[0,\infty)$, we can determine the ratio of two functional determinants of the operators with mass $m$, using the Gel'fand and Yaglom's theorem \cite{gy}, as
\begin{equation}
\frac{\det({\cal M} +m^2) }{\det({\cal M}^{\rm free} +m^2)} =  \frac{\psi(\infty)}{\psi^{\rm free}(\infty)},
\end{equation}
where the wave function $\psi(r)$ satisfies the ordinary differential equation (ODE) $({\cal  M}+m^2)\psi=0$ with initial
value conditions at $r=0$: $\psi(0)=0$ and $\psi'(0)=1$. The other function $\psi^{\rm free}(r)$ is the solution to the
differential equation $({\cal  M}^{\rm free}+m^2)\psi^{\rm free}=0$ with the same initial conditions. This method turns
the problem of finding an infinite number of eigenvalues into that of finding the solutions to the ODE initial value problems.

Now we consider the case of higher dimensions (i.e., $d\geq 2$). [In this paper we will provide explicit formulas for the cases with $d=2,3,4,5$ but the extension to higher dimension is also straightforward]. When the potential is radial, i.e., $V=V(r)$, we can use the partial wave analysis, taking advantage of the spherical symmetry. Formally, for the radially separable operators given in (\ref{op}), the logarithm of the determinant ratio can be written as a sum of the logarithm of radial (that is, one-dimensional) determinant ratios:
\begin{eqnarray}
\Omega = \ln\left( \frac{\det[{\cal M}+m^2]}{\det[{\cal M}^{\rm free}+m^2]} \right) = \sum_{l=0}^\infty g_l(d) \ln\left( \frac{\det[{\cal M}_l+m^2]}{\det[{\cal M}_l^{\rm free}+m^2]}  \right).
\label{detlsum}\end{eqnarray}
Here $l$ denotes the angular momentum quantum number appropriate to each partial wave and
\begin{equation}
 g_l(d)=\frac{(2l+d-2)(l+d-3)!}{l!(d-2)!}
\end{equation}
is the degeneracy factor \cite{GR, dunnekirsten}. The associated radial differential operator${\cal M}_l$ is given by
\begin{eqnarray}
{\cal M}_l &=& -\frac{1}{r^{d-1}}\frac{\partial}{\partial r}\left(r^{d-1} \frac{\partial}{\partial r}\right) +\frac{l(l+d-2)}{r^2} +V(r),
\end{eqnarray}
and ${\cal M}_l^{\rm free}$ has the same form as ${\cal M}_l$ but without the potential term $V(r)$.

The individual radial determinant ratio in (\ref{detlsum}) is finite and it can be evaluated easily  by using the above Gel'fand-Yaglom method; but the sum over $l$ leads to a divergent result. This problem is related to renormalization and an elegant method to extract the finite or renormalized expression $\Omega_{\rm ren}$ from $\Omega$ (after a suitable regularization and renormalization) is presented in \cite{idet, dunnekirsten}. We are concerned here with more practical problem, which should be addressed if one wants the full, including the finite part, expression of $\Omega_{\rm ren}$. The rate of convergence of the $l$-sum in (\ref{detlsum}) is quite slow, and we require an efficient method to deal with this $l$-sum. To this end it is convenient to introduce a partial wave cutoff $L$ \cite{radial1} and to split the sum into two pieces, i.e., the low angular momentum part $\Omega_{\rm L}$ and the high angular momentum part $\Omega_{\rm H}$:
\begin{eqnarray}
&& \Omega_{\rm ren} = \Omega_{\rm L} +\Omega_{\rm H},    \label{lcut}  \\
&& \Omega_{\rm L} = \sum_{l=0}^L g_l(d) \ln\left( \frac{\det[{\cal M}_l+m^2]}{\det[{\cal M}_l^{\rm free}+m^2]} \right), \label{lcut1}\\
&& \Omega_{\rm H} = \sum_{l=L+1}^\infty g_l(d) \ln\left( \frac{\det[{\cal M}_l+m^2]}{\det[{\cal M}_l^{\rm free}+m^2]} \right)+ \delta\Omega, \label{lcut2}
\end{eqnarray}
where $\delta\Omega$ denotes the `conventional' renormalization counterterm. Separate treatment of $\Omega_{\rm L}$ and $\Omega_{\rm H}$ constitute the crux of the partial wave cutoff method.

The part $\Omega_{\rm L}$ (see (\ref{lcut1})) can be evaluated with the help of the Gel'fand-Yaglom method. Since the determinant ratio for given $l$ behaves like $\sim \frac{1}{l}$ for large $l$ and the degeneracy factor $g_l(d)$ increases as $l^{d-2}$, it should be clear that $\Omega_{\rm L}$ behaves like $L^{d-2}$ for $d\geq3$ and like $\ln L$  for $d=2$  in the large $L$ limit. (This reveals the divergent structures in the formal expression in (\ref{detlsum})). As for the part $\Omega_{\rm H}$ which involves the sum of all partial wave contributions with $l\geq L+1$, we can evaluate it {\em analytically} in a uniform asymptotic series of the form
\begin{equation}
\Omega_{\rm H}=\int_0^\infty dr \left( Q_{\rm log} +\sum_{n=2-d}^\infty Q_{-n}L^{-n}\right), \label{asympL}
\end{equation}
where $Q_{-n}$'s may have an implicit $L$ dependency of $O(L^0)$ and $Q_{\rm log}$ behaves as $O(\ln L)$ in the large $L$ limit. This uniform nature makes also the $r$ integrals in (\ref{asympL}) well-defined. To find explicit forms of the $Q$'s, we take the proper-time representation for the functional determinant of radial operators for each partial wave and then use the quantum mechanical radial-WKB expansion which becomes exact in the large $l$ limit. For the desired large $L$ expansion we then perform the sum over $l=L+1, \cdots, \infty$ with the help of the Euler-Maclaurin method. These are given in following sections.

Note that, as $L\to\infty$, unsuppressed terms in the expansion (\ref{asympL}) may grow like $\ln L, L, \cdots, L^{(d-2)}$, but they match precisely the large-$L$ divergences coming from $\Omega_{\rm L}$ except for the sign. Hence, combining the large-$L$-unsuppressed terms of $\Omega_{\rm H}$ with $\Omega_{\rm L}$ and taking $L\to\infty$ limit, we get a finite renormalized quantity $\Omega_{\rm ren}$, i.e.,
\begin{equation}
\Omega_{\rm ren}= \lim_{L\to\infty}\left[\Omega_{\rm L} + \int_0^\infty dr
\left( Q_{\rm log}  +\sum_{n=0}^{d-2} Q_{n}L^{n}\right)\right]. \label{subtracted}
\end{equation}
In \cite{dunnekirsten}, Dunne and Kirsten identified this expression for $d=2,3,4$ by using the zeta function technique.

In principle, since (\ref{subtracted}) yields a well-defined expression, one can use this expression to obtain the renormalized functional determinant. But we still have a practical problem determining $\Omega_{\rm L}$. Since it is generally not possible to find a master formula for the determinant ratio valid for all $l$, we have to evaluate (numerically) those partial wave contributions corresponding to the angular momentum range $0\leq l\leq L$, to be able to determine $\Omega_{\rm L}$. Because of the slow rate of convergence, a very large number of these determinant terms should be thus considered to get a sufficiently good result for the sum. There is a rather simple way to secure a reliable large-$L$ limit value in (\ref{subtracted}) with a relatively small number of partial wave contributions. Including the $\frac{1}{L}$-suppressed terms of the expansion (\ref{asympL}) inside the squared braces in (\ref{subtracted}) would make the sum converge faster, thereby reducing the number of partial-wave determinants to be evaluated explicitly. To appreciate this better, note that if we were able to calculate both of $\Omega_{\rm L}$ and $\Omega_{\rm H}$ {\em exactly}, their sum would be independent of the choice of the cutoff $L$.
%But the expression in the squared braces (\ref{subtracted}) gets an $L$-dependency and it disappears only in the limit $L\to\infty$.
This implies that, leaving aside possible numerical inaccuracy in calculating $\Omega_{\rm L}$, the $L$-dependency in the sum for finite value of the cutoff $L$ is really due to our ignoring of the $\frac{1}{L}$-suppressed contributions in the asymptotic series (\ref{asympL}). Therefore, it is possible to reduce this undesired $L$-dependency systematically by taking into account the (ignored) higher order terms in the $\frac{1}{L}$-series. Now, instead of taking the strict $L\to\infty$ limit in (\ref{subtracted}), we can write the following formula for $\Omega_{\rm ren}$:
\begin{equation}
\Omega_{\rm ren}= \Omega_{\rm L} + \int_0^\infty dr \left( Q_{\rm log}  +\sum_{n=0}^{d-2} Q_{n}L^{n} +\sum_{n=1}^{N} Q_{-n}\frac{1}{L^{n}} \right) +O\left(\frac{1}{L^{N+1}}\right),  \label{truncated}
\end{equation}
where $N$ refers to the order of truncation. In this formula the error is indicated by the last term and it is totally under control. It is apparent that, for a given value of the cutoff $L$, we get more accurate value of $\Omega_{\rm ren}$ by taking into account more $\frac{1}{L}$-suppressed terms. Or, for a given accuracy, we can lower the value of $L$ by including some $\frac{1}{L}$-suppressed terms. We can thus use (\ref{truncated}) as the basis of precision calculation for functional determinants. In this work, we take $N=4$ for concreteness and derive the expressions for $Q_{d-2}, \cdots, Q_{-4}$ and $Q_{\rm log}$ in dimensions $d=2,3,4,5$ to facilitate the use of our formula (\ref{truncated}) in various physical problems.

\section{The Proper-time Radial Green Function and Its Derivative Expansion} \label{wkbsection}
First we write the partial-wave determinant ratio in a more convenient form
\begin{equation}
\frac{\det({\cal M}_l+ m^2)}{\det({\cal M}_l^{\rm free}+ m^2)}
=\frac{\det(\tilde{\cal M}_l+ m^2)}{\det(\tilde{\cal M}_l^{\rm free}+ m^2)},
\end{equation}
where
\begin{eqnarray}
\tilde{\cal M}_l &\equiv & r^{(d-1)/2}{\cal M}_l r^{-(d-1)/2} \\
&=& - \frac{d^2}{dr^2} +\frac{(l+\frac{d-3}{2}) (l+\frac{d-1}{2})}{r^2} + V(r)
\end{eqnarray}
and $\tilde{\cal M}_l^{\rm free}$ is equal to $\tilde{\cal M}_l$ with $V=0$. Note that the operators $\tilde{\cal M}_l$ and $\tilde{\cal M}_l^{\rm free}$ do not involve any first order derivative term. It is also convenient to introduce the effective potential
\begin{equation}
{\cal V}_l(r)=\frac{(l+\frac{d-3}{2}) (l+\frac{d-1}{2})}{r^2} + V(r).
\end{equation}

The Schwinger proper-time representation for the determinant ratio, for a partial wave $l$, is given as
\begin{equation}
\frac{\det(\tilde{\cal M}_l+ m^2)}{\det(\tilde{\cal M}_l^{\rm free}+ m^2)}
=- \int_0^\infty \frac{ds}{s}\; e^{-m^2s} \int_0^\infty dr
 \left\{ \Delta_l(r,r;s) -\Delta_l^{\rm free}(r,r;s)\right\}.
\end{equation}
Here the proper-time radial Green function is defined by
\begin{eqnarray}
\Delta_l(r,r';s) = \langle r| e^{-s\tilde {\cal M}_l} |r' \rangle
\label{radialGreendef}
\end{eqnarray}
($\Delta_l^{\rm free}$ corresponds to the one with $\tilde{\cal M}_l^{\rm free}$ instead of $\tilde{\cal M}_l$) and it satisfies the equation
\begin{equation}
\left\{ \partial_s -\partial^2_r + {\cal V}_l(r)\right\} \Delta_l(r,r';s)=0.
\label{pwSE}\end{equation}
Since we are interested in the large $l$ behavior of the Green function, let us rescale the potential ${\cal V}_l$ and the proper-time $s$, following \cite{radial1}, as
\begin{equation}
{\cal V}_l=l^2 {\cal U}_l, \qquad s=\frac{t}{l^2}.
\end{equation}
Then (\ref{pwSE}) becomes
\begin{equation}
\left\{ \partial_t -\frac{1}{l^2}\partial^2_r + {\cal U}_l(r)\right\} \Delta_l\left(r,r';\frac{t}{l^2}\right)=0.
\label{pwSE2}\end{equation}
From this equation one may readily recognize that the large $l$ expansion is of the same nature as the derivative expansion which is an expansion in the number of derivatives on ${\cal V}_l$.

In the rest of this section, we develop the derivative expansion of the above proper-time radial Green function.
The resulting series will be identical with the large $l$ series considered in \cite{radial1}, but a simpler derivation is given here.  Introducing the momentum variable $p$, the proper-time Green function can be cast into the form
\begin{eqnarray}
\Delta(r,r';s) &=& \int_{-\infty}^\infty \frac{dp}{2\pi} \langle r| e^{-s\tilde{\cal M}_l} |p \rangle \langle p|r' \rangle \nonumber\\
&=& \int_{-\infty}^\infty \frac{dp}{2\pi} e^{-s [ -\partial_r^2 + {\cal V}_l(r)]} e^{-ip\;(r-r')} .
\end{eqnarray}
After moving the last Fourier factor $e^{-ip(r-r')}$ to the left of the differential operator $\partial_r$, one may set the coincident limit $r'=r$ and get
\begin{eqnarray}
\Delta(r,r;s) &=& \int_{-\infty}^\infty \frac{dp}{2\pi}
e^{-s [ -( \partial_r-ip)^2 +{\cal V}_l (r)]}\nonumber\\
&\equiv& \int_{-\infty}^\infty \frac{dp}{2\pi} e^{ -sp^2} K(r,p;s), \label{GreenK}
\end{eqnarray}
where a new  function $K(r,p;s)$ is introduced.  $K(r,p;s)$ satisfies the following differential equation
\begin{eqnarray}
\left\{ {\partial_s} -{\partial_r^2} +2ip {\partial_r} +{\cal V}_l(r) \right\} K(r,p;s) = 0
\label{diffeqK}\end{eqnarray}
and the boundary condition $K(r,p;0)=1$.

Now, we introduce an auxiliary expansion parameter $\lambda$ in (\ref{diffeqK}), i.e., consider
\begin{eqnarray}
\left( {\partial_s} -\lambda^2 {\partial_r^2} +2i \lambda p{\partial_r} +{\cal V}_l(r) \right\} K(r,p;s) = 0. \label{Kequation}
\end{eqnarray}
Then, taking $\lambda$ as an expansion parameter, a series solution to the above equation can be found with the ansatz:
\begin{eqnarray}
K(r,p;s) = e^{-s {\cal V}_l(r)} \left[1 +\lambda b_1(r,p;s) +\lambda^2 b_2(r,p;s) +\cdots\right]. \label{Kseries}
\end{eqnarray}
Plugging (\ref{Kseries}) into (\ref{Kequation}), we find the recurrence relations
\begin{eqnarray}
\partial_s b_k &=& -2ip \left\{ {\partial_r} -s {\cal V}_l' \right\} b_{k-1} +\left\{ \partial_r^2  - 2s {\cal V}_l' \partial_r -s {\cal V}_l''+ s^2 ({\cal V}_l')^2 \right\} b_{k-2}, \quad (k\geq2)
\label{recrel}\end{eqnarray}
together with $b_0=1$ and $\partial_s b_1 = 2ips {\cal V}_l'(r)$. With the boundary conditions $b_k(r,p;0)=0$, these recurrence relations may be used to determine the coefficient functions $b_k$ ($k=1,2,\cdots$). Some leading terms satisfying (\ref{recrel}) are easily found. Note that, $b_k(r,p;s)$ is a simple odd/even polynomial of $p$ when $k$ is odd/even and hence it vanishes as we integrate over $p$ for all odd numbers of $k$. Here we report a few leading $b_k$'s  with even numbers of $k$ ($k=2,4,6$):
\begin{eqnarray}
&& b_2 = \left(\frac{s^3}{3}-\frac{p^2 s^4}{2}\right) ({\cal V}_l')^2+\left(\frac{2 p^2 s^3}{3}-\frac{s^2}{2}\right) {\cal V}_l'', \\
&& b_4 = \left(\frac{p^4 s^8}{24}-\frac{p^2 s^7}{6}+\frac{s^6}{18}\right) ({\cal V}_l')^4+\left(-\frac{p^4 s^7}{3}+\frac{47 p^2 s^6}{36}-\frac{13 s^5}{30}\right) {\cal V}_l'' ({\cal V}_l')^2  \nonumber\\
&& \,\, +\left(\frac{p^4 s^6}{3}-\frac{19 p^2 s^5}{15}+\frac{5 s^4}{12}\right) {\cal V}_l^{(3)} {\cal V}_l'+\left(\frac{2 p^4 s^6}{9}-\frac{13 p^2 s^5}{15}+\frac{7 s^4}{24}\right) ({\cal V}_l'')^2 \nonumber\\
&& \,\, +\left(-\frac{2 p^4 s^5}{15}+\frac{p^2 s^4}{2}-\frac{s^3}{6}\right) {\cal V}_l^{(4)}, \\
&& b_6 = \left(\frac{p^6 s^{11}}{36}-\frac{13 p^4 s^{10}}{48}+\frac{287 p^2 s^9}{540}-\frac{7 s^8}{60}\right) {\cal V}_l'' ({\cal V}_l')^4+\left(\frac{4 p^6 s^7}{315}-\frac{p^4 s^6}{9}+\frac{p^2 s^5}{5}-\frac{s^4}{24}\right) {\cal V}_l^{(6)} \nonumber\\
&& \,\, -\left(\frac{2 p^6 s^8}{45}-\frac{124 p^4 s^7}{315}+\frac{43 p^2 s^6}{60}-\frac{3 s^5}{20}\right) {\cal V}_l^{(5)} {\cal V}_l'-\left(\frac{p^6 s^{12}}{720}-\frac{p^4 s^{11}}{72}+\frac{p^2 s^{10}}{36}-\frac{s^9}{162}\right) ({\cal V}_l')^6 \nonumber\\
&& \,\, -\left(\frac{p^6 s^8}{18}-\frac{32 p^4 s^7}{63}+\frac{17 p^2 s^6}{18} -\frac{s^5}{5}\right) ({\cal V}_l^{(3)})^2+\left(\frac{4 p^6 s^9}{81}-\frac{7 p^4 s^8}{15}+\frac{25 p^2 s^7}{28}-\frac{139 s^6}{720}\right) ({\cal V}_l'')^3 \nonumber\\
&& \,\, -\left(\frac{p^6 s^{10}}{9}-\frac{287 p^4 s^9}{270}+\frac{493 p^2 s^8}{240}+\frac{25 s^7}{56}\right) ({\cal V}_l'')^2 ({\cal V}_l')^2-\left(\frac{4 p^6 s^8}{45}-\frac{254 p^4 s^7}{315}+\frac{269 p^2 s^6}{180} \right. \nonumber\\
&& \quad \left. -\frac{19 s^5}{60}\right) {\cal V}_l'' {\cal V}_l^{(4)} +\left(\frac{p^6 s^9}{15}-\frac{109 p^4 s^8}{180}+\frac{473 p^2 s^7}{420}-\frac{43 s^6}{180}\right) {\cal V}_l^{(4)} ({\cal V}_l')^2+\left(\frac{2 p^6 s^9}{9}-\frac{31 p^4 s^8}{15} \right. \nonumber\\
&& \quad \left. +\frac{82 p^2 s^7}{21}-\frac{301 s^6}{360}\right) {\cal V}_l'' {\cal V}_l^{(3)} {\cal V}_l'-\left(\frac{p^6 s^{10}}{18}-\frac{47 p^4 s^9}{90}+\frac{359 p^2 s^8}{360}-\frac{271 s^7}{1260}\right) {\cal V}_l^{(3)} ({\cal V}_l')^3.
\end{eqnarray}

From the series solution of $K(r,p;s)$ we perform the Gaussian integrations over $p$. Then the proper time radial Green function at the same points, $\Delta(r,r;s)$, is determined as
\begin{eqnarray}
&& \Delta(r,r;s) = \frac{e^{-s {\cal V}_l}} {\sqrt{4\pi s}} \left[ 1 +\lambda^2 \left( \frac{s^3}{12}({\cal V}_l')^2 -\frac{s^2}{6}{\cal V}_l'' \right) \right. \nonumber\\
&& \qquad +\lambda^4 \left( \frac{s^6}{288} ({\cal V}_l')^4 -\frac{11s^5}{360} ({\cal V}_l')^2 {\cal V}_l'' + \frac{s^4}{40} ({\cal V}_l'')^2 +\frac{s^4}{30} {\cal V}_l' {\cal V}_l^{(3)} -\frac{s^3}{60} {\cal V}_l^{(4)} \right) \nonumber\\
&& \qquad +\lambda^6 \left(\frac{ s^9}{10368}({\cal V}_l')^6-\frac{17  s^8}{8640}({\cal V}_l')^4 {\cal V}_l'' +\frac{83  s^7}{10080}({\cal V}_l'{\cal V}_l'')^2 +\frac{s^7}{252} ({\cal V}_l')^3 {\cal V}_l^{(3)} -\frac{61s^6}{15120} ({\cal V}_l'')^3 \right.   \nonumber\\
&& \qquad\qquad -\frac{43s^6}{2520}  {\cal V}_l' {\cal V}_l'' {\cal V}_l^{(3)}
-\frac{5s^6}{1008}({\cal V}_l')^2 {\cal V}_l^{(4)} +\frac{23s^5}{5040} ({\cal V}_l^{(3)})^2 +\frac{19s^5}{2520} {\cal V}_l''  {\cal V}_l^{(4)} \nonumber\\
&& \qquad\qquad \left. \left. +\frac{s^5}{280} {\cal V}_l' {\cal V}_l^{(5)} -\frac{s^4}{840} {\cal V}_l^{(6)} \right) +O\left(\lambda^8\right) \right] . \label{Greenseries}
\end{eqnarray}
Clearly, this $\lambda$-series is organized according to the total number of derivative on ${\cal V}_l$. Of course, we can set $\lambda=1$ now. We also remark that, ignoring the last sixth order term in (\ref{Greenseries}), the quantum mechanical WKB series  used in the approximate evaluation of the instanton determinant in \cite{idetpl} can be obtained from (\ref{Greenseries}) if we substitute the relevant expression for the potential $V(r)$.

\section{Large $L$ Expansion of the High Angular Momentum Part} \label{radialsection}
In this section we use the derivative expansion (\ref{Greenseries}) to derive the large $L$ expansion in (\ref{asympL}). To this end it is necessary to identify the structure of the renormalization counterterms, $\delta\Omega$, first. For simplicity we use the dimensional renormalization method. Setting the dimension of space-time to be $d-2\epsilon$, a dimensionally regularized expression of the functional determinant is, using the proper-time representation,
\begin{eqnarray}
\Omega_{\epsilon} = -\int_0^\infty \frac{ds}{s} \frac{e^{-m^2s}\mu^{2\epsilon}}{(4\pi)^{d/2} s^{d/2-\epsilon}} \int d^d{\bf x} \sum_{k=1}^\infty a_k({\bf x},{\bf x})s^k, \label{dimregularization}
\end{eqnarray}
where $\mu$, introduced for dimensional reason, carries the dimension of mass. In (\ref{dimregularization}), $a_k({\bf x},{\bf x})$ ($k=1,2,3,\cdots$) are the heat-kernel coefficients.  Even if many of them are known in explicit forms, two leading coefficients, $a_1=-V $ and $a_2=\frac{1}{2}V^2-\frac{1}{6}\partial^2V $, are sufficient for our purpose. The integration over $s$ yields
\begin{eqnarray}
\Omega_\epsilon = -\frac{m^d}{(4\pi)^{d/2}} \left( \frac{\mu}{m} \right)^{2\epsilon} \sum_{k=1}^\infty \frac{\Gamma( k-\frac{d}{2}+\epsilon)}{m^{2k}} \int d^d{\bf x}\; a_k({\bf x},{\bf x}).
\end{eqnarray}
When the dimension of space is an odd number, above expression is finite at $\epsilon=0$ and does not require any counterterm. However it has a pole when the associated dimension is even and,  particularly for $d=2,4$, it has the structure:
\begin{eqnarray}
\Omega_\epsilon &\sim& -\frac{1}{\epsilon}\int \frac{d^2{\bf x}}{4\pi} a_1({\bf x},{\bf x}), \qquad (d=2), \\
\Omega_\epsilon &\sim& -\frac{1}{ \epsilon} \int \frac{d^4{\bf x}}{16\pi^2} [-m^2a_1({\bf x},{\bf x})+a_2({\bf x},{\bf x})], \quad (d=4),
\end{eqnarray}
in the $\epsilon\to 0$ limit. For $d=2$ and $4$ we here choose the renormalization counterterms, assuming the minimal subtraction scheme, as follows:
\begin{eqnarray}
\delta \Omega &=& \frac{1}{4\pi} \left( \frac{1}{\epsilon} -\gamma_E \right) \int d^2{\bf x}\; a_1({\bf x},{\bf x}), \qquad (d=2), \\
\delta\Omega &=& \frac{1}{16\pi^2} \left( \frac{1}{\epsilon} -\gamma_E \right) \int d^4{\bf x} \left[ -m^2a_1({\bf x},{\bf x})+a_2({\bf x},{\bf x}) \right],  \quad (d=4),
\end{eqnarray}
where $\gamma_E$ is Euler's constant. With these counterterms, $\Omega_{\rm H}$ has a finite expression
\begin{eqnarray}
\Omega_{\rm H} = \lim_{\epsilon\to 0}\left[ -\int_0^\infty \frac{ds}{s}\; (\mu^2s)^\epsilon e^{-m^2s} \int_0^\infty dr \sum_{l=L+1}^\infty g_l(d) \left\{ \Delta_l(r,r;s) -\Delta_l^{\rm free}(r,r;s) \right\} +\delta\Omega \right]. \label{largecontributionrenormalized}
\end{eqnarray}

From the expression (\ref{largecontributionrenormalized}) and the derivative expansion of the proper-time radial Green function $\Delta_l(r,r;s)$ obtained in the previous section, the desired large $L$ expansion in (\ref{asympL}) can be derived. After plugging (\ref{Greenseries}) into (\ref{largecontributionrenormalized}), we may perform the $l$ summation first.
Note that the sum has the structure
\begin{eqnarray}
\sum_{l=L+1}^\infty e^{-s (A_2l^2+A_1l+A_0)} (c_0+c_1l+c_2l^2+\cdots).
\label{sumstructure}
\end{eqnarray}
This kind of summation cannot be done explicitly; but, with the help of Euler-Maclaurin method, it is possible to obtain the large-$L$ series expansion. In computing the asymptotic expansion of this sum, the most useful form of Euler-Maclaurin formula is
\begin{eqnarray}
\sum_{l=L+1}^\infty f(l) = \int_L^\infty f(l)dl -\frac{1}{2}f(L) -\frac{1}{12}f'(L) +\frac{1}{720}f^{(3)}(L) +\cdots,
\end{eqnarray}
assuming that $f(l)$ and its derivatives vanish at $l\to\infty$. When $f(l)$ is of the form (\ref{sumstructure}), the integral in Euler-Maclaurin formula can be performed in terms of the exponential function and the error function along with a polynomial of $L$. Regarding $s$ to be of an order of $1/L^2$,
we find that taking a derivative of $f(L)$ increases the power of $1/L$. Therefore the rest of the series in Euler-Maclaurin formula is an asymptotic large-$L$ series. See Appendix C in \cite{radial1} for more details. After using Euler-Maclaurin formula and changing the integration variable $s$ with $\frac{t}{L^2}$, we can find the large $L$ series for the high angular momentum contribution in the following form:
\begin{eqnarray}
\sum_{l=L+1}^\infty g_l(d) \left\{ \Delta_l(r,r;s) -\Delta_l^{\rm free}(r,r;s) \right\} = \sum_{n=2-d}^\infty P_{-n} L^{-n}. \label{Pseries}
\end{eqnarray}
Because of the presence of the degeneracy factor, explicit forms of $P_{-n}$'s and further evaluation depend on the dimension of the space. However, since the procedure itself is basically the same, we will present the related calculation in detail for $d=2$ and only the final results for $d=3,4,5$ below.

\subsection{2D}
Note that the degeneracy factor is simply $g_l(2)=2$, which is independent from $l$. The large $L$ expansion in (\ref{Pseries}) starts from $P_0$.
Some of the leading coefficient functions $P_{-n}$ are explicitly evaluated as
\begin{eqnarray}
P_0 &=& -\frac{r}{2} {\rm erfc}\left[ \frac{\sqrt{t}}{r} \right] V, \\
P_{-1} &=& \frac{e^{-\frac{t}{r^2}}}{2\sqrt{\pi}} \sqrt{t}V, \\
P_{-2} &=& \frac{e^{-\frac{t}{r^2}}}{6\sqrt{\pi}} \left\{ \frac{t^{3/2}}{2r^2} (V-2rV') -\frac{t^{5/2}}{r^4} V \right\} +\frac{{\rm erfc}\left[ \frac{\sqrt{t}}{r} \right]}{12} t(3rV^2-V'-rV'').
\end{eqnarray}
Other coefficient functions have similar structures: one part being $e^{-{t}/{r^2}}$ times a polynomial in $V$ or its derivatives, and the other part $\mathrm{erfc}(\frac{\sqrt{t}}{r})$ times a different polynomial in $V$ or its derivatives.
One may perform $t$ integration with these explicit forms of $P_{-n}$'s.  In performing this process with the first term, $P_0$, we find a pole term: explicitly,
\begin{eqnarray}
-\int_0^\infty \frac{dt}{t} e^{ -\frac{m^2t}{L^2}}  \left( \frac{\mu^2t}{L^2} \right)^\epsilon {\rm erfc}\left( \frac{\sqrt{t}}{r} \right)
= -\frac{1}{\epsilon} +\gamma_E -2\ln\left(\frac{\mu r}{(1+u)L}\right)  + O(\epsilon)
\end{eqnarray}
with
\begin{equation}
u=\sqrt{1+\frac{m^2r^2}{L^2}} .
\end{equation}
This $\frac{1}{\epsilon}$ divergence is canceled by the renormalization counterterm $\delta\Omega$.
In the evaluation of other terms on the other hand, no such divergence arises and so the limit $\epsilon\to 0$ can be taken safely. In the evaluation of these terms, following integral formulas are useful:
\begin{eqnarray}
\int_0^\infty \frac{dt}{t} t^n  e^{ -t\frac{u^2}{r^2} } &=& \frac{r^{2n}}{u^{2n}} \Gamma(n), \label{gausint} \\
\int_0^\infty \frac{dt}{t} t^n e^{ -\frac{m^2t}{L^2}} {\rm erfc}\left( \frac{\sqrt{t}}{r} \right) &=& 2\left( \frac{r}{2} \right)^{2n} \frac{\Gamma(2n)}{\Gamma(n+1)}\;  _2F_1(n,n+1/2; n+1; 1-u^2), \label{erfcint}
\end{eqnarray}
where $_2F_1$ is the hypergeometric function. Here note that we do not expand the function
$u = \sqrt{ 1+ \frac{m^2 r^2 }{L^2} }$
as a power series of $\frac{m^2r^2}{L^2}$ (even for a large value of $L$), for this kind of expansion breaks down when $m$ or/and $r$ get large.  Keeping the function $u$ as a whole, we can maintain the uniform nature of our large $L$ expansion.

After the $l$-sum and the $t$ integration, we can generate a $\frac{1}{L}$ series for the large partial wave contribution $\Omega_{\rm H}$ as desired. The leading term of this series is
\begin{eqnarray}
&& Q_{\rm log} = \ln\left( \frac{\mu r}{(u+1)L} \right) r V, \label{q2log}
\end{eqnarray}
which is the only nonvanishing term as we let $L\to\infty$ (since $Q_0=0$).  Other terms vanish in the limit $L\to\infty$ like $L^{-n}$ ($n\geq 1$), but they can be important for $L$ not too large. Some of those secondary leading terms, needed for the fast evaluation of $\Omega_{\rm ren}$, are found to have following forms:
\begin{eqnarray}
&& Q_{-1} = -\frac{1}{2u} r V, \\
&& Q_{-2} = \frac{1}{24u^5(u+1)} \left\{ -6 r^3 u^4 V^2+2 r^3 u^4 V''+2 r^2 (u^2+u+1) u^2 V' \right. \nonumber\\
&& \qquad \left. -r (u^3+u^2-3 u-3) V \right\}, \\
&& Q_{-3} = \frac{1}{48u^7} \left\{ 6 r^3 u^4 V^2-2 r^3 u^4 V''-6 r^2 u^2 V'-3 r (u^4-6 u^2+5) V \right\}, \\
&& Q_{-4} = \frac{1}{1920 L^4 u^{11} (u+1)^2} \left\{ 80 r^5 (2 u+1) u^8 V^3-40 r^5 (2 u+1) u^8 (V')^2 \right. \nonumber\\
&& \qquad +8 r^5 (2 u+1) u^8 V^{(4)}+16 r^4 (2 u^3+4 u^2+6 u+3) u^6 V^{(3)}+60 r^3 (u+1)^2 (u^2-5) u^4 V^2 \nonumber\\
&& \qquad -r (u+1)^2 (81 u^6-1185 u^4+2695 u^2-1575) V -80 r^4 (2 u^3+4 u^2+6 u+3) u^6 V V' \nonumber\\
&& \qquad +4 r^2 (4 u^7+18 u^6+32 u^5-139 u^4-310 u^3+20 u^2+350 u+175) u^2 V' \nonumber\\
&& \qquad \left. -4 r^3 (4 u^5+13 u^4+22 u^3-44 u^2-110 u-55) u^4 V''-80 r^5 (2 u+1) u^8 V V'' \right\}.
\end{eqnarray}

The formula given in \cite{dunnekirsten} can be reproduced immediately, utilizing only the piece $Q_{\rm log}$ above.
First note that their result in 2D can be written in the form
\begin{eqnarray}
\Omega_{\rm DK} = \lim_{L\to\infty} \left[ \sum_{l=0}^L 2 \Omega_l  +\int_0^\infty dr r V \left\{ -\sum_{l=1}^L \frac{1}{l} +\ln\left(\frac{\mu r}{2}\right) +\gamma_E \right\} \right].
\label{2dDK}\end{eqnarray}
Now, using the relation
\begin{eqnarray}
-\sum_{l=1}^L \frac{1}{l} +\ln\left(\frac{\mu r}{2}\right) +\gamma_E = \ln\left(\frac{\mu r}{2 L}\right) +O(L^{-1}),
\end{eqnarray}
one can see that the second part of (\ref{2dDK}) indeed corresponds to $Q_{\rm log}$. (Here observe that $u\to 1$ as $L\to \infty$). This clearly shows that the result of Dunne and Kirsten, $\Omega_{\rm DK}$, is equal to
\begin{equation}
\Omega_{\rm DK}= \lim_{L\to\infty}\left[  \sum_{l=0}^L 2 \Omega_l + \int_0^\infty dr
Q_{\rm log}\right],
\end{equation}
i.e., our expression without any $\frac{1}{L}$-suppressed terms involving $Q_{-1},Q_{-2},\cdots$.

\subsection{3D}
In 3 dimensions no ultraviolet divergence problem arises in the dimensional regularization procedure and hence
no renormalization counterterm is necessary, i.e., $\delta\Omega=0$. Now the degeneracy factor is $g_l(3)=2l+1$ -- it grows linearly with $l$. Hence the series in (\ref{Pseries}) starts from $P_1$. Using the integral formulas in
(\ref{gausint}) and (\ref{erfcint}), we can perform the $t$ integration explicitly to obtain following expressions for the $Q$'s:
\begin{eqnarray}
&& Q_1 = -u r V, \\
&& Q_0 = -\frac{1}{u} r V, \\
&& Q_{-1} = \frac{1}{24u^5} \left\{ -6 r^3 u^4 V^2+2 r^3 u^4 V''+2 r^2 (2 u^2+1) u^2 V'-3 r (4 u^4-3 u^2-1) V \right\}, \\
&& Q_{-2} = \frac{1}{24u^7} \left\{ 6 r^3 u^4 V^2-2 r^3 u^4 V''-6 r^2 u^2 V'+3 r (2 u^4+3 u^2-5) V \right\}, \\
&& Q_{-3} = \frac{1}{1920u^{11}} \left\{ 80 r^5 u^8 V^3+60 r^3 (4 u^4-9
u^2-5) u^4 V^2+8 r^5 u^8 V^{(4)}-40 r^5 u^8 (V')^2 \right. \nonumber\\
&& \,\, +16 r^4 (2 u^2+3) u^6 V^{(3)}-20 r^3 (4 u^4-9 u^2-11) u^4 V''-20 r^2 (30 u^4-19 u^2-35) u^2 V' \nonumber\\
&& \,\, \left. -80 r^5 u^8 V V''-80 r^4 (2 u^2+3) u^6 V V'+15 r (88 u^6-235 u^4+42 u^2+105) V \right\}, \\
&& Q_{-4} = \frac{1}{1920u^{13}} \left\{ 120 r^5 u^8 (V')^2-60 r^3 (6 u^4+15 u^2-35) u^4 V^2-24 r^5 u^8 V^{(4)}-240 r^5 u^8 V^3 \right. \nonumber\\
&& \,\, -240 r^4 u^6 V^{(3)}+20 r^3 (6 u^4+39 u^2-77) u^4 V''-60 r^2 (4 u^6+18 u^4-119 u^2+105) u^2 V' \nonumber\\
&& \,\, \left. +240 r^5 u^8 V V''+1200 r^4 u^6 V V'+15 r (48 u^8-20 u^6-1015 u^4+2142 u^2-1155) V \right\}.
\end{eqnarray}
These explicit results can be utilized for fast evaluations of $\Omega_{\rm ren}$ in $d=3$. Note that there is no $\ln L$ related term in this dimension.

If one does not care much about the fast convergence of the expression (in the limit $L\to\infty$), the renormalized quantity $\Omega_{\rm ren}$ can be found using only the terms $Q_1$ and $Q_0$ above, i.e.,
\begin{equation}
\Omega_{\rm ren}= \lim_{L\to\infty}\left[ \sum_{l=0}^L (2l+1) \Omega_l  + \int_0^\infty dr \left(Q_{1}L +Q_0\right)\right].
\label{3dren}\end{equation}
On the other hand, the 3D formula for the same quantity given in \cite{dunnekirsten} reads
\begin{eqnarray}
%\Omega_{\rm ren} &=& \sum_{l=0}^\infty (2l+1) \left\{ \ln\left( \frac{\det[{\cal M}_l+m^2]}{\det[{\cal M}_l^{\rm free}+m^2]} \right) -\frac{\int_0^\infty dr r V(r)}{2l+1}\right\} \nonumber\\
\Omega_{\rm DK}= \lim_{L\to\infty} \left[ \sum_{l=0}^L (2l+1) \Omega_l -\int_0^\infty dr r V(r) (L+1) \right].
\label{3dDK}\end{eqnarray}
Observing that $u=1+O(L^{-2})$, one may easily see that
\begin{eqnarray}
(Q_1 L +Q_0)= - r V(r) (L+1)  +O(L^{-1})
\end{eqnarray}
and therefore two quantities in (\ref{3dDK}) and (\ref{3dren}) are identical.

\subsection{4D}\label{4d}
The degeneracy factor in 4 dimensions is $g_l(4)=(l+1)^2$ -- it grows like a quadratic power in $l$. The large $L$ series in (\ref{Pseries}) starts from $P_2$ in this case.  The integral formulas in (\ref{gausint}) and (\ref{erfcint}) enable us to perform the $t$ integration again. After some calculations we have found the following results for the unsuppressed quantities:
\begin{eqnarray}
&& Q_{\rm log} = -\frac{1}{24} \ln\left( \frac{\mu r}{(u+1)L} \right) (6 m^2 r^3 V+3 r^3 V^2-3 r^2 V'-r^3 V''), \label{4dqlog}\\
&& Q_2 = -\frac{1}{8} (u^2+2u-1) r V, \\
&& Q_1 = -\frac{3}{4u} r V, \\
&& Q_0 = \frac{1}{48 u^5} \left\{ -6 r^3 u^4 V^2+2 r^3 u^4 V''+2 r^2 (3 u^2+1) u^2 V'+r (-52 u^4+25 u^2+3) V \right\}. \qquad \label{4dq0}
\end{eqnarray}
Subleading terms in the $\frac{1}{L}$ asymptotic expansion can also be identified, with the results
\begin{eqnarray}
&& Q_{-1} = \frac{1}{32 u^7} \left\{ 6 r^3 u^4 V^2-2 r^3 u^4 V''-6 r^2 u^2 V'-r (16 u^6-37 u^4+6 u^2+15) V \right\}, \\
&&Q_{-2} = \frac{1}{11520 u^{11} (u^2-1)} \left\{ 240 r^5 (u^3-1) u^8 V^3-120 r^5 (u^3-1) u^8 (V')^2 \right. \nonumber\\
&& \quad +24 r^5 (u^3-1) u^8 V^{(4)}+144 r^4 (u^5-1) u^6 V^{(3)}+60 r^3 (52 u^6-127 u^4+60 u^2+15) u^4 V^2 \nonumber\\
&& \quad +4 r^3 (18 u^7-260 u^6+635 u^4-228 u^2-165) u^4 V''-720 r^4 (u^5-1) u^6 V V' \nonumber\\
&& \quad -12 r^2 (6 u^9+510 u^6-1011 u^4+320 u^2+175) u^2 V'-240 r^5 (u^3-1) u^8 V V'' \nonumber\\
&& \quad \left. +3 r (u^2-1)^2 (3288 u^6+1605 u^4-7630 u^2-1575) V \right\}, \\
&& Q_{-3} = \frac{1}{7680 u^{13}} \left\{ -720 r^5 u^8 V^3+360 r^5 u^8 (V')^2-72 r^5 u^8 V^{(4)}-720 r^4 u^6 V^{(3)} \right. \nonumber\\
&& \quad +60 r^3 (16 u^6-111 u^4+30 u^2+105) u^4 V^2-20 r^3 (16 u^6-111 u^4-42 u^2+231) u^4 V'' \nonumber\\
&& \quad -60 r^2 (90 u^6-201 u^4-182 u^2+315) u^2 V'+720 r^5 u^8 V V''+3600 r^4 u^6 V V' \nonumber\\
&& \quad \left. +15 r (96 u^{10}+63 u^8-2980 u^6+3010 u^4+3276 u^2-3465) V \right\}, \\
&& Q_{-4} = \frac{1}{645120 u^{17} (u^2-1)^2} \left\{ r^7 (2 u^5-5 u^2+3) u^{12} \left[ -1680 V^4 +3360 V (V')^2 -1008 (V'')^2 \right. \right. \nonumber\\
&& \qquad \left. +3360 V^2 V'' -1344 V' V^{(3)} -672 V V^{(4)} +48 V^{(6)}\right] 
  +r^6 (2 u^7-7 u^2+5) u^{10} \left[432 V^{(5)} \right. \nonumber\\
&& \qquad \left. -7392 V' V'' -4032 V V^{(3)} +10080 V^2 V' \right]
 -1680 r^5 (u^2-1)^2 (52 u^4-125 u^2-35) u^8 V^3 \nonumber\\
&& \quad +5 r (u^2-1)^3 ( 64692 u^{10}-748223 u^8-1201788 u^6+7638246 u^4-4708704 u^2 \nonumber\\
&& \qquad  -1576575 )  V -168 r^5 (28 u^9-260 u^8+1145 u^6-1083 u^4-355 u^2+525)  u^8 (V')^2 \nonumber\\
&& \quad -336 r^5 (12 u^9-260 u^8+1145 u^6-1167 u^4-115 u^2+385) u^8 V V'' \nonumber\\
&& \quad +24 r^5 (36 u^9-364 u^8+1603 u^6-1365 u^4-785 u^2+875) u^8 V^{(4)} \nonumber\\
&& \quad +1008 r^4 (4 u^{11}+850 u^8-2869 u^6+2645 u^4-105 u^2-525) u^6 V V' \nonumber\\
&& \quad -144 r^4 (8 u^{11}+1190 u^8-3843 u^6+3030 u^4+665 u^2-1050) u^6 V^{(3)} \nonumber\\
&& \quad -42 r^3 (u^2-1)^2 (9864 u^8-8415 u^6-64645 u^4+54495 u^2+17325) u^4 V^2 \nonumber\\
&& \quad +2 r^3 (432 u^{13}+69048 u^{12}-25641 u^{10}-1242409 u^8+2941010 u^6-1987510 u^4 \nonumber\\
&& \qquad -167265 u^2+412335) u^4 V''-6 r^2 (144 u^{15}+46032 u^{14}-228564 u^{12}-659897 u^{10} \nonumber\\
&& \qquad \left. +4287475 u^8-6809390 u^6+3736110 u^4+153615 u^2-525525) u^2 V' \right\}.
\end{eqnarray}

These can be used for fast evaluations of $\Omega_{\rm ren}$ in $d=4$. For the comparison with the result of \cite{dunnekirsten}, we here give the formula presented in \cite{dunnekirsten}:
\begin{eqnarray}
\Omega_{\rm DK} &=& \lim_{L\to\infty} \left[ \sum_{l=0}^L (l+1)^2 \left\{ \Omega_l -\frac{\int_0^\infty dr r V(r)}{2(l+1)} +\frac{\int_0^\infty dr r^3 V(V+2m^2)}{8(l+1)^3} \right\} \right. \nonumber\\
&& \qquad\left. -\frac{1}{8} \int_0^\infty dr r^3 V(V+2m^2) \left\{ \ln\left(\frac{\mu r}{2}\right) +\gamma_E +1 \right\} \right].
\label{4dDK}\end{eqnarray}
The $l$ sum of the second term in the above equation yields
\begin{eqnarray}
\sum_{l=0}^L (l+1)^2 \left\{ -\frac{r V(r)}{2(l+1)} \right\} = -r V(r) \left( \frac{L^2}{4} +\frac{3L}{4} +\frac{1}{2} \right),
\end{eqnarray}
while the $l$ sum of the next term, if combined with the last term, produces
\begin{eqnarray}
&& \frac{V(V+2m^2)}{8} \left\{ \sum_{l=0}^L \frac{1}{l+1} -\ln\left( \frac{\mu r}{2} \right) -\gamma_E -1 \right\} \nonumber\\
&& \qquad = - \frac{V(V+2m^2)}{8} \left\{ \ln\left( \frac{\mu r}{2L} \right) +1 \right\} +O(L^{-1}).
\end{eqnarray}
On the other hand, our formula for $\Omega_{\rm ren}$ using only unsuppressed terms (given in (\ref{4dqlog})-(\ref{4dq0})) is
\begin{eqnarray}
&& \Omega_{\rm ren} = \lim_{L\to\infty} \left[ \sum_{l=0}^L (l+1)^2 \Omega_l +\int_0^\infty dr (Q_{\rm log} +Q_2 L^2 +Q_1 L +Q_0) \right], \\
&& Q_{\rm log} +Q_2 L^2 +Q_1 L +Q_0 \nonumber\\
&& \qquad= \ln\left( \frac{\mu r}{2L} \right) \left\{ -\frac{r^3V(V+2m^2)}{8} +\frac{(r^3V')'}{24}  \right\} -r V(r) \left( \frac{L^2}{4} +\frac{3L}{4} +\frac{1}{2} \right) \nonumber\\
&& \qquad\qquad -\frac{r^3V(V+2m^2)}{8} +\frac{1}{24} (4r^2 V' +r^3 V'' ) +O(L^{-1}).
\label{q0123}\end{eqnarray}
The difference between $\Omega_{\rm ren}$ and (\ref{4dDK}) is
\begin{eqnarray}
\Omega_{\rm ren} -\Omega_{\rm DK}= \int_0^\infty dr \frac{1}{24} \left\{ r^3 V' \ln\left( \frac{\mu r}{2L} \right) +r^3 V' \right\}'=0,
\end{eqnarray}
showing that our result is consistent with the 4D formula given in \cite{dunnekirsten}.

\subsection{5D}
The degeneracy factor in the 5 dimensional space is $g_l(5)=\frac{1}{6}(l+1)(l+2)(2l+3)$. The first term in the large $L$ series in (\ref{Pseries}) will be the term $P_3$ now. The integral formulas in (\ref{gausint}) and (\ref{erfcint}) are used in the $t$ integration again. After the $l$ summation and the $t$ integrations, we have found the following results:
\begin{eqnarray}
&& Q_3 = \frac{1}{18} u(2u^2-3) r V, \\
&& Q_2 = -\frac{1}{3u} r V, \\
&& Q_1 = \frac{1}{144u^5} \left\{ 6 r^3 (2 u^2-1) u^4 V^2+2 r^3 (1-2 u^2) u^4 V''+2 r^2 (-8 u^4+4 u^2+1) u^2 V' \right. \nonumber\\
&& \quad \left. +r (-138 u^4+47 u^2+3) V \right\}, \\
&& Q_0 = \frac{1}{72u^7} \left\{ 6 r^3 u^4 V^2-2 r^3 u^4 V''-6 r^2 u^2 V'-3 r (28 u^6-34 u^4+9 u^2+5) V \right\}, \\
&& Q_{-1} = \frac{1}{11520 u^{11}} \left\{ 80 r^5 (2 u^2+1) u^8 V^3-40 r^5 (2 u^2+1) u^8 (V')^2+8 r^5 (2 u^2+1) u^8 V^{(4)} \right. \nonumber\\
&& \quad +16 r^4 (8 u^4+4 u^2+3) u^6 V^{(3)}+60 r^3 (46 u^4-47 u^2-5) u^4 V^2 \nonumber\\
&& \quad -5 r (1152 u^{10}-5600 u^8+5010 u^6+2469 u^4-2716 u^2-315) V \nonumber\\
&& \quad -80 r^4 (8 u^4+4 u^2+3) u^6 V V'+4 r^3 (32 u^6-214 u^4+247 u^2+55) u^4 V'' \nonumber\\
&& \quad \left. -4 r^2 (32 u^8+16 u^6+1172 u^4-1045 u^2-175) u^2 V'-80 r^5 (2 u^2+1) u^8 V V'' \right\}, \\
&& Q_{-2} = \frac{1}{5760 u^{13}} \left\{ -240 r^5 u^8 V^3+120 r^5 u^8 (V')^2-24 r^5 u^8 V^(4)-240 r^4 u^6 V^{(3)} \right. \nonumber\\
&& \quad +60 r^3 (28 u^6-102 u^4+45 u^2+35) u^4 V^2-20 r^3 (28 u^6-102 u^4+21 u^2+77) u^4 V'' \nonumber\\
&& \quad -180 r^2 (32 u^6-74 u^4+7 u^2+35) u^2 V'+240 r^5 u^8 V V''+1200 r^4 u^6 V V' \nonumber\\
&& \quad \left. +15 r (704 u^{10}-1656 u^8-1260 u^6+3745 u^4-378 u^2-1155) V \right\}.
\end{eqnarray}
One may use these results for fast evaluation of $\Omega_{\rm ren}$ in $d=5$.

\section{The Prefactor in False Vacuum Decay Rate}
We will here illustrate how the formulas obtained in previous sections can be used to improve the rate of convergence in the calculation of the prefactor in the false vacuum decay. We consider a simple four-dimensional scalar field theory described by the Euclidean action
\begin{equation}
S_{\rm cl}[\phi]=\int d^4 x\left\{ \frac{1}{2}(\partial_\mu \phi)^2+U(\phi)\right\}\quad ,
\label{action}
\end{equation}
with
\begin{equation}
U(\phi)=\frac{\lambda}{8}\left(\phi^2-a^2\right)^2-\frac{\epsilon}{2 a}\left(\phi-a\right)\quad .
\label{pot1}
\end{equation}
Here the parameter $\epsilon$ represents a constant external source, which serves to break the degeneracy of the double-well-type potential. The potential $U(\phi)$ has two nondegenerate classical minima, $\phi_-$ and $\phi_+(>\phi_-)$, with $U(\phi_-) > U(\phi_+)$.
After expanding the field $\phi$ about the false vacuum $\phi_-$
\begin{equation}
\phi=\phi_- + \varphi  ,
\label{expansion}
\end{equation}
it is convenient to rescale the field $\varphi$ and the spacetime coordinates as
\begin{equation}
\bar{x}= m x\quad ;  \quad\varphi =\frac{m^2}{2\eta} \Phi
\label{rescale}
\end{equation}
in the dimensionless form.  Here the parameters $m$ and $\eta$  are related to the original couplings by
\begin{equation}
m^2=\frac{\lambda}{2}\left(3\phi_-^2-a^2\right)\quad ; \quad \eta=\frac{\lambda}{2}|\phi_- |.
\label{meta}
\end{equation}
Then the classical action in terms of these dimensionless quantities is
\begin{equation}
S_{\rm cl}[\Phi]=\left(\frac{m^2}{4\eta^2}\right) \int d^4 \bar{x}\left[\frac{1}{2}(\bar{\partial}_\mu \Phi)^2+\frac{1}{2}\Phi^2-\frac{1}{2}\Phi^3+\frac{\alpha}{8}\Phi^4\right]\quad
\label{action2}
\end{equation}
with the dimensionless quartic coupling constant $\alpha=\frac{\lambda m^2}{4 \eta^2}$.
%For our semiclassical analysis, we assume that the overall factor is greater than 1, i.e.,$\frac{m^2}{4\eta^2}\gg 1$.

The bounce $\Phi_{\rm cl}(r)$, which determines the decay of false vacuum, is a solution to the nonlinear ordinary differential equation
\begin{equation}
-\Phi_{\rm cl}^{\prime\prime} -\frac{3}{r}\Phi_{\rm cl}^\prime +\Phi_{\rm cl}-\frac{3}{2}\Phi_{\rm cl}^2+\frac{\alpha}{2} \Phi_{\rm cl}^3=0,
\label{bounceeq}
\end{equation}
satisfying the boundary conditions
\begin{eqnarray}
\Phi_{\rm cl}^\prime(0)=0, \qquad
\Phi_{\rm cl}(\infty)=0 .
\label{bcb}\end{eqnarray}
It is hard to solve this equation analytically but one can always find a numerical solution.
The false vacuum decay rate is denoted by $\gamma_{\rm decay}$ and in the one-loop approximation it is given by \cite{coleman}
\begin{equation}
\gamma_{\rm decay}= \left(\frac{S_{\rm cl}[\Phi_{\rm cl}]}{2\pi}\right)^2 \left(\frac{\det' \left[-\partial^2 + U^{\prime\prime}(\Phi_{\rm cl})\right]}{\det \left[-\partial^2 +U^{\prime\prime}(\Phi_-)\right]}\right)^{-1/2}\,
e^{-S[\Phi_{\rm cl}]-\delta\Omega}\quad ,
\label{rate}
\end{equation}
where the prime on the determinant means that the zero modes (corresponding to translational moves) are removed.  In the exponent the first term $S[\Phi_{\rm cl}]$ denotes the classical action of the bounce and  $\delta \Omega$ denotes the renormalization counterterms. Note that this quantity $\gamma_{\rm decay}$ involves the functional determinant and thus it can be evaluated using the methods developed in this paper.  The potential $V(r)$ in (\ref{op}) is now fixed as
\begin{equation}
V(r)= U^{\prime\prime}(\Phi_{\rm cl}) =-3\Phi_{\rm cl}(r)+\frac{3\alpha}{2}\Phi_{\rm cl}^2(r),
\label{radialpot}
\end{equation}
and so the effective potential ${\cal V}_l$ in the radial operator ${\cal M}_l$ for a partial wave with $l$ becomes
\begin{equation}
{\cal V}_l(r)=\frac{(l+\frac{1}{2})(l+\frac{3}{2})}{r^2}-3\Phi_{\rm cl}(r)+\frac{3\alpha}{2}\Phi_{\rm cl}^2(r) .
\label{effpot}
\end{equation}

For the comparison with the result of \cite{dunnemin}, we will call the logarithm of the prefactor in $\gamma_{\rm decay}$ (with the opposite sign) the effective action $\Gamma$. Then the partial wave expression for the renormalized effective action is
\begin{eqnarray}
\Gamma &=&\frac{1}{2}|\Omega_0| + \frac{1}{2}\sum_{l=2}^{L}(l+1)^2 \Omega_l -2 \ln \left[\frac{\pi}{2} \Phi_\infty \left(\Phi_0-\frac{3}{2}\Phi_0^2+\frac{\alpha}{2}\Phi_0^3\right)\right]\nonumber  \\
&&+ \int dr \left(Q_{\rm log}+\sum_{n=-2}^{4} Q_{-n} \frac{1}{L^n} \right) +O\left(\frac{1}{L^5}\right)
\label{effectiveaction}
\end{eqnarray}
with $\Omega_l= \ln \det (\tilde{\cal M}_l +m^2)/\ln\det (\tilde{\cal M}_l^{\rm free} +m^2)$.
In this expression, $\Omega_0$ has a negative sign and its absolute value is taken. The contribution from the sector with $l=1$ involves the zero modes related to translational invariance and it is, having been removed from the sum, written down separately.  The analytic expression for that contribution has been found in \cite{dunnemin}.
The factor $\frac{1}{2}$ in front of (\ref{effectiveaction}) is introduced since we are considering a real single scalar field
and $(l+1)^2$ denotes the degeneracy factor. In the last term,
$\Phi_0\equiv\Phi_{\rm cl}(0)$ and $\Phi_\infty$ is the coefficient of $K_1(r)/r$  ($K_1(r)$ denotes the modified Bessel function) in the large $r$ behavior of $\Phi_{\rm cl}(r)$.  The coefficient functions $Q$'s are explicitly given in our subsection \ref{4d}.

\begin{figure}
\includegraphics[scale=1.2]{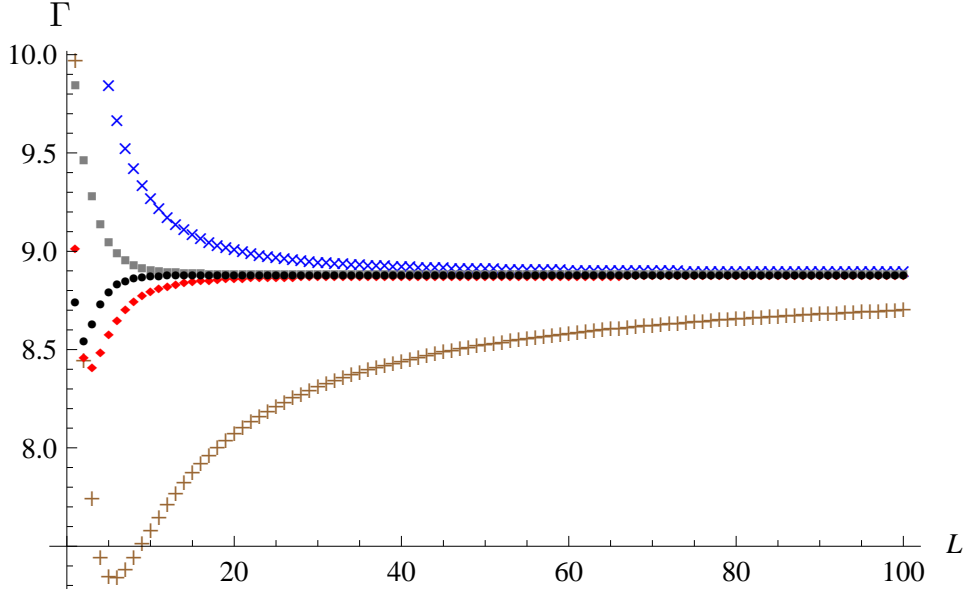}
\caption{Plot for the sum of the low angular momentum ($l\leq L$) and  the high momentum part when we take $L=1,2,\cdots,100$.  The (brown) pluses denote the values as we ignore all terms of $O(\frac{1}{L})$ and beyond. Slow convergence is evident. Solving 100 differential equations to determine the determinant for each partial wave (for the case with $L=100$) is not enough to approach the limit. The blue crosses denote those after incorporating $O(\frac{1}{L})$ corrections and show the marked improvement of convergence already. The (red) diamonds, (gray) squares, and (black) dots represent the cases obtained after we incorporate $\frac{1}{L^2}$, $\frac{1}{L^3}$, and $\frac{1}{L^4}$ corrections successively.}
\label{fig1}
\end{figure}

\begin{figure}
\includegraphics[scale=1.2]{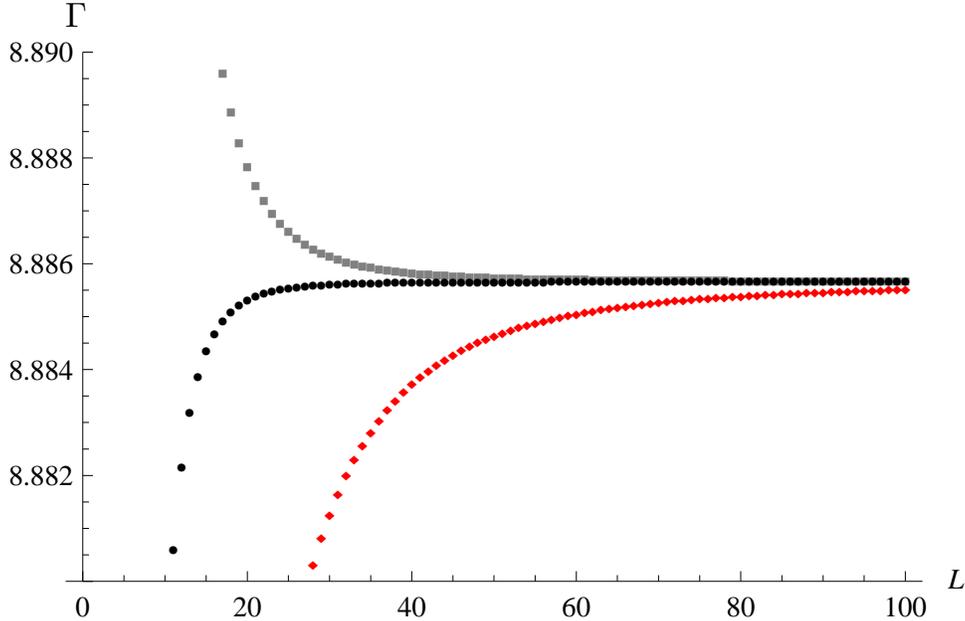}
\caption{The same plots with Fig \ref{fig1} but the scale is  magnified by 250. We can see only the (red) diamonds, (gray)squares, and (black) dots which represent the values after incorporating $\frac{1}{L^2}$, $\frac{1}{L^3}$, and $\frac{1}{L^4}$ corrections successively. One may conclude that $L\sim 30$ is enough to get the limiting value with a very good precision.}
\label{fig2}
\end{figure}

In the case $\alpha=0.5$ we plot in FIG. \ref{fig1} the numerical values for the right hand side of (\ref{effectiveaction}) as a function of $L=1,2,\cdots,100$, taking $\frac{1}{L}$-suppressed parts of the asymptotic expansion successively.
The lowest plots in this figure is the case when we ignore all suppressed terms in the large-$L$ expansion of the high angular momentum part. It shows the existence of the $L\to\infty$ limit. But, since the rate of convergence is quite slow, it is difficult to find the actual limit value which is the desired value for the effective action $\Gamma$. Other plots represent the values obtained after incorporating $\frac{1}{L^2}$, $\frac{1}{L^3}$, and $\frac{1}{L^4}$ corrections successively and they show remarkable improvements in convergence as we include these corrections.  Magnified form of this figure is shown in the FIG \ref{fig2}. A relatively small number of $L$, for instance $L\sim30$ in the $\alpha=0.5$ case, produces a good convergence and thus gives us a very accurate number of the effective action when $\int dr [\sum_{n=1}^{4} Q_{-n} L^{-n}]$ is added.
\newpage
\section{Radial Operators in Gauge Theories} \label{gaugesection}
In this section we consider an application of the large $L$ expansion in (\ref{asympL}) to gauge theory. In \cite{radial2} it was used to calculate the renormalized effective actions in classes of radially symmetric background gauge fields.
We present here the explicit forms of some coefficients functions $Q_{-n}$ which were announced in that paper.

The one-loop effective action in a gauge theory is expressed by the logarithm of the functional determinant,
\begin{eqnarray}
\Omega = \ln\left( \frac{\det[-D^2+m^2]}{\det[-\partial^2+m^2]} \right),
\end{eqnarray}
where $D^2 = D_\mu D_\mu$ and $D_\mu = \partial_\mu-iA_\mu$ is the covariant derivative operator.  We consider two cases with SU(2) backgrounds gauge fields of the form
\begin{eqnarray}
&\text{(Case 1): }& A_{\mu}({\bf x}) = 2 \eta_{\mu\nu a} x_\nu f(r) \frac{\tau^a}{2}, \label{case1} \\
&\text{(Case 2): }& A_{\mu}({\bf x}) = 2 (\eta_{\mu\nu i} \hat{u}_i) x_\nu g(r) \frac{\tau^3}{2}, \label{case2}
\end{eqnarray}
where $\mu,\nu = 1,2,3,4$, %$a=1,2,3$, $\tau$'s denote $2\times2$ Pauli matrices,
$\eta_{\mu\nu a}$ (or $\eta_{\mu\nu i}$) are the 't Hooft symbols \cite{thooft} and $\hat{u}_i$ a unit 3-vector. The functions, $f(r)$ and $g(r)$ are unspecified so that general (radially symmetric) background fields can be studied.  Following \cite{thooft}, we define the angular momentum operators $L_a \equiv -\frac{i}{2} \eta_{\mu\nu a} x_\mu \partial_\nu$ and the SU(2) isospin generators $T_a = \frac{\tau^a}{2}$, which satisfy the commutation relations $[L_a,L_b] = i \epsilon_{abc} L_c$ and $[T_a,T_b] = i \epsilon_{abc} T_c$.
These operators carry the quantum numbers characterized by  $l$ ($=0,\frac{1}{2},1,\cdots$) and $t$ ($=\frac{1}{2}$).
Note that the angular momentum operators defined here are different from those of previous sections so that the quantum number $l$ can take half-integer values as well as integer values. (The quantum number $l$ in previous sections corresponds to $2l$ in this section). Because of the radial symmetry, as in previous sections, the one-loop effective action can be written as a sum over one-dimensional radial determinants.

\subsection{Case 1}
The form of the gauge field (\ref{case1}) is inspired by the instanton solution %\cite{instanton}
(which corresponds to $f(r)=\frac{1}{r^2+\rho^2}$ with the size parameter $\rho$) and it carries genuine non-Abelian nature. The covariant Laplacian $-D^2$ involves the isospin-orbit coupling term and then $\vec J^2$ ($J_a \equiv L_a +T_a$), $\vec L^2$ and $\vec T^2=\frac{3}{4}$ are conserved quantities.  Therefore, partial waves are specified by the quantum numbers $(l,j)$. 
For each sector, the radial differential operator $\tilde{\cal M}_{(l,j)}$ associated with the covariant Laplacian $-D^2$ assumes the form
\begin{equation}
\tilde{\cal M}_{(l,j)} = -\partial_r^2 + {\cal V}_{(l,j)}
\end{equation}
with the effective potential
\begin{equation}
{\cal V}_{(l,j)}= \frac{(2l+\frac{1}{2})(2l+\frac{3}{2})}{r^2}
 +4f(r) \left\{ j(j+1)-l(l+1)-\frac{3}{4} \right\} +3r^2f(r)^2.
\end{equation}
The corresponding radial operator for the free Laplacian $-\partial^2$ is
\begin{eqnarray}
\tilde{\cal M}_l^{\rm free} = -\partial_r^2 + {\cal V}_l^{\rm free}= -\partial_r^2 +\frac{(2l+\frac{1}{2})(2l+\frac{3}{2})}{r^2}.
\end{eqnarray}

Introducing the one-dimensional radial determinant $\Omega_{(l,j)}$
\begin{eqnarray}
\Omega_{(l,j)} = \ln \left( \frac{\det[\tilde{\cal M}_{(l,j)}+m^2]}{\det[\tilde{\cal M}_l^{\rm free}+m^2]} \right),
\end{eqnarray}
the low angular momentum part of the one-loop effective action can be written as
\begin{eqnarray}
\Omega_{\rm L} = \sum_{l=0,\frac{1}{2},1,\cdots}^L (2l+1)(2l+2)\left\{ \Omega_{(l,l+\frac{1}{2})} +\Omega_{(l+\frac{1}{2},l)} \right\}
\label{case1L}\end{eqnarray}
and the corresponding high angular momentum part as
\begin{eqnarray}
\Omega_{\rm H} = \sum_{l=L+\frac{1}{2}}^\infty (2l+1)(2l+2)\left\{ \Omega_{(l,l+\frac{1}{2})} +\Omega_{(l+\frac{1}{2},l)} \right\}+
\delta \Omega \label{case1H}
\end{eqnarray}
with the renormalization counterterm
\begin{equation}
\delta\Omega=\left(\frac{1}{\epsilon}-\gamma_E\right)\int_0^\infty dr \frac{r^3}{8} \left[ 4r^4f(r)^4-8r^2f(r)^3+8f(r)^2+4rf'(r)f(r)+r^2f'(r)^2 \right].
\end{equation}
[The Pauli-Villas regularization was employed in \cite{radial2} but we have changed it to the dimensional regularization scheme in this work]. In (\ref{case1L}) and (\ref{case1H}) we have rearranged $j(=l\pm\frac{1}{2})$-sum by combining two sectors of $(l,j=l+\frac{1}{2})$ and $(l+\frac{1}{2},j=l)$ with the same degeneracy factor $(2l+1)(2l+2)$. See \cite{radial1} for details.

Using the WKB series described in Sec.\ref{wkbsection}, the high partial-wave contribution (\ref{case1H}) can be calculated analytically in the form of large-$L$ asymptotic series. The calculational step is almost the same as that described in Sec. \ref{radialsection}. A part of the result was presented in \cite{radial2} by the form 
\begin{eqnarray}
\Omega_{\rm H} &=& \int_0^\infty dr \left\{ Q_{\rm log} \ln L +\sum_{n=-2}^\infty Q_{-n}L^{-n} \right\}. \label{largecontributiongauge}
\end{eqnarray}
The first few terms, i.e., $Q_{\rm log}$, $Q_2$, $Q_{1}$, $Q_{0}$ and $Q_{-1}$ are given by relatively short expressions and they are already presented in \cite{radial2}.  We  here report the explicit expressions for $Q_{-2}$ and $Q_{-3}$:
\begin{eqnarray}
&& Q_{-2} = \frac{1}{7680 r u^{13} (u^2-1)} \left\{ -80 (16 u^9 +27 u^6 -81 u^4 + 45 u^2 - 7) u^4 G^3 \right. \nonumber\\
&& \quad +20 (32 u^{11} +792 u^{10} -5175 u^8 +8910 u^6 -5140 u^4 +266 u^2 +315) u^2 G^2 \nonumber\\
&& \quad +15 (u^2 -1)^2 (3472 u^8 -5619 u^6 -7399 u^4 +7623 u^2 +1155) G \nonumber\\
&& \quad -40 r (12 u^9 +54 u^6 -171 u^4 +140 u^2 -35) u^4 (G^2)' -60 r^2 (4 u^7 +6 u^4 -15 u^2 +5) u^6 (G')^2 \nonumber\\
&& \quad +40 r^2 (4 u^7 +9 u^4 -18 u^2 +5) u^6 (G^2)'' +48 r^3 u^6 (u^2 -1) (4 u^2 -5) G^{(3)} \nonumber\\
&& \quad +4 r (1760 u^{10} -12494 u^8 +25779 u^6 -19595 u^4 +2975 u^2 +1575) u^2 G' \nonumber\\
&& \quad -4 r^2 (440 u^8 -1889 u^6 +2019 u^4 -185 u^2 -385) u^4 G'' -24 r^4 (u^2 -1) u^8 G^{(4)} \nonumber\\
&& \quad +24 r^4 (u^5-1) u^8 (3 (H'')^2 +4 H' H^{(3)}) -16 r^3 (u^7 -36 u^2 +35) u^6 H' H'' \nonumber\\
&& \quad \left. +4 r^2 (6 u^9 -930 u^6 +1854 u^4 -685 u^2 -245) u^4 (H')^2 \right\}, \\
&& Q_{-3} = \frac{1}{7680 r u^{15}} \left\{ -720 (9 u^6 -45 u^4 +35 u^2 -7) u^4 G^3 -60 r^2 (18 u^4 -75 u^2 +35) u^6 (G')^2 \right. \nonumber\\
&& \quad +60 (72 u^{10} -1431 u^8 +3170 u^6 -420 u^4 -2562 u^2 +1155) u^2 G^2 \nonumber\\
&& \quad +15 (528 u^{12} -3833 u^{10} -8245 u^8 +41454 u^6 -29442 u^4 -15477 u^2 +15015) G \nonumber\\
&& \quad -40 r u^4 (162 u^6 -855 u^4 +980 u^2 -315) (G^2)' +40 r^2 u^6 (27 u^4 -90 u^2 +35) (G^2)'' \nonumber\\
&& \quad +12 r (160 u^{10} -3534 u^8 +10165 u^6 -4655 u^4 -7875 u^2 +5775) u^2 G' \nonumber\\
&& \quad -12 r^2 (40 u^8 -449 u^6 +165 u^4 +1435 u^2 -1155) u^4 G'' +48 r^3 u^6 (12 u^4 -45 u^2 +35) G^{(3)} \nonumber\\
&& \quad -24 r^4 (3 u^2 -5) u^8 G^{(4)} -120 r^4 u^8 (3(H'')^2 +4 H' H^{(3)}) +80 r^3 u^6 (36 u^2 -49) H' H'' \nonumber\\
&& \quad \left. -20 r^2 (174 u^6 -414 u^4 -161 u^2 +441) u^4 (H')^2 \right\},
\end{eqnarray}
where $u(r)=\sqrt{1+\frac{m^2r^2}{4L^2}}$, $H(r)=r^2 f(r)$ and $G(r)=H(r)(H(r)-1)$. The $Q_{-4}$ term was also used for the evaluation in \cite{radial2}, but the expression for $Q_{-4}$ is quite long and its actual numerical value is rather small in most cases. So we do not present it here.

In \cite{radial2} it was shown that incorporating the combination $\sum_{n=1}^4 Q_{-n}L^{-n}$ made the summation over $l$ converge dramatically fast. For instance, FIG. 4b in \cite{radial2} clearly demonstrated the changes when each of $Q_{-1}$, \ldots, $Q_{-4}$ terms was added one by one.

\subsection{Case 2}
The second case is quasi-Abelian. The field has a fixed color direction and only $\{L^2, L_3, T_3\}$ are conserved quantities.
Partial waves for the covariant Laplacian $-D^2$ are classified with the quantum numbers $(l,l_3,t_3)$ (for $l_3=-l,\cdots,l$, $t_3={ \pm \frac{1}{2}}$). The radial operator for a given partial wave has the form:
\begin{eqnarray}
\tilde{\cal M}_{(l,l_3,t_3)} &=& -\partial_r^2 +{\cal V}_{(l,l_3,t_3)}, \\
{\cal V}_{(l,l_3,t_3)}&=&  \frac{(2l+\frac{1}{2})(2l+\frac{3}{2})}{r^2}+8g(r)t_3 l_3 +r^2g(r)^2.
\end{eqnarray}
The one-loop effective action in this case can be written as
\begin{eqnarray}
\Omega = \sum_{l=0,\frac{1}{2},1,\cdots}^L \sum_{l_3=-l}^l \sum_{t_3=\pm\frac{1}{2}} (2l+1)\Omega_{(l,l_3,t_3)} +\sum_{l=L+\frac{1}{2}}^\infty \sum_{l_3=-l}^l \sum_{t_3=\pm\frac{1}{2}} (2l+1)\Omega_{(l,l_3,t_3)}, \label{case2split}
\end{eqnarray}
where $\Omega_{(l,l_3,t_3)}$ is the one-dimensional radial determinant defined by
\begin{eqnarray}
\Omega_{(l,l_3,t_3)} = \ln \left( \frac{\det[-D^2_{(l,l_3,t_3)}+m^2]}{\det[-\partial_{(l)}+m^2]} \right).
\end{eqnarray}

As in Case 1, the result of renormalized large partial wave contribution can be written in the form (\ref{largecontributiongauge}) and $Q_2,\cdots,Q_{-1}$ as well as $Q_{\rm log}$ terms were given in \cite{radial2}. The expressions for $Q_{-2}$ and $Q_{-3}$ are
\begin{eqnarray}
&& Q_{-2} = \frac{r^3}{23040 u^{13} (u^2-1)} \left\{ -240 u^4 (u^2-1)^3 r^8 g^6+240 u^6 (u^2-1) (4 u^2-3) r^6 g^2 (g')^2 \right. \nonumber\\
&& \quad +480 u^6 (u^2-1)^2 r^6 g^3 g''+480 u^4 (u^2-1) (6 u^4+3 u^2-7) r^5 g^3 g'-48 g u^8 (u^2-1) r^4 g^{(4)} \nonumber\\
&& \quad +60 u^2 (52 u^{10}-303 u^8+678 u^6-560 u^4+70 u^2+63) r^4 g^4 \nonumber\\
&& \quad +24 u^8 (u^5-2 u^2+1) r^4 (3(g'')^2+4g' g^{(3)})+16 u^6 (71 u^7-72 u^4-54 u^2+55) g' g'' r^3 \nonumber\\
&& \quad +4 u^4 (442 u^9-520 u^8+1606 u^6-2043 u^4-10 u^2+525) r^2 (g')^2 \nonumber\\
&& \quad +8 u^4 (216 u^9-260 u^8+1043 u^6-1599 u^4+215 u^2+385) r^2 g g'' \nonumber\\
&& \quad +40 u^2 (72 u^11-208 u^{10}+202 u^8+1131 u^6-1855 u^4+343 u^2+315) r g g' \nonumber\\
&& \quad +5 (2248 u^{12}-12421 u^{10}+11733 u^8+20070 u^6-32214 u^4+7119 u^2+3465) g^2 \nonumber\\
&& \quad \left. +96 u^6 (2 u^7-4 u^4-3 u^2+5) r^3 g g^{(3)} \right\}, \\
&& Q_{-3} = \frac{r^3}{30720 u^{15}} \left\{ -720 u^4 (u^2-3) (u^2-1)^2 r^8 g^6+240 u^6 (12 u^4-35 u^2+21) r^6 g^2 (g')^2 \right. \nonumber\\
&& \quad +480 u^6 (3 u^4-10 u^2+7) r^6 g^3 g''+480 u^4 (18 u^6-15 u^4-70 u^2+63) r^5 g^3 g' \nonumber\\
&& \quad -48 u^8 (3 u^2-5) r^4 g g^{(4)}+60 u^2 (16 u^{10}-207 u^8+960 u^6-1022 u^4-504 u^2+693) r^4 g^4 \nonumber\\
&& \quad -24 u^8 (6 u^2-5) r^4 (3(g'')^2+4g' g^{(3)})-16 u^6 (216 u^4+270 u^2-385) r^3 g' g'' \nonumber\\
&& \quad -96 u^6 (12 u^4+15 u^2-35) r^3 g g^{(3)}-8 u^4 (80 u^8-699 u^6+2595 u^4+1645 u^2-3465) r^2 g g'' \nonumber\\
&& \quad -4 u^4 (160 u^8-1038 u^6+3465 u^4+3220 u^2-4725) r^2 (g')^2 \nonumber\\
&& \quad -40 u^2 (64 u^{10}-174 u^8-1605 u^6+2905 u^4+2583 u^2-3465) r g g' \nonumber\\
&& \quad \left. +5 (160 u^{12}-4107 u^{10}-735 u^8+49770 u^6-44982 u^4-46431 u^2+45045) g^2 \right\}.
\end{eqnarray}
Since the radial operator explicitly depends on $l_3$, we have to calculate the radial determinant for each $l_3$: this makes the amount of calculation grow very fast, i.e., by quadratic powers of $L$ as $L$ becomes large. Thus the effect of acceleration, rendered by incorporating the $\frac{1}{L}$-suppressed terms, is greater than other cases.

\section{Conclusion}
We have here presented an efficient and precise method for the calculation of functional determinants with radially symmetric differential operators.
This method involves the partial wave cutoff technique in which the infinite partial wave summation is separated into two sectors as in (\ref{lcut}).  The first sector is evaluated with the radial Gel'fand-Yaglom method for each partial wave.
We have developed the large-$L$ asymptotic series for the second sector, i.e., for the high angular momentum part.
Combining the first sector with the unsuppressed terms of the latter series,  the renormalized sum can be found in the $L\to\infty$ limit. Including the subsequent (i.e., $\frac{1}{L}$-suppressed) terms in the series also, we can get a precise value for the functional determinant with the choice of relatively small $L$ (which means less computational work for low angular momentum part). Certainly this greatly improves the efficiency of calculation. That is, we can get an result with estimated errors of  $\sim\frac{1}{L}$ if we ignore all the terms suppressed by $\frac{1}{L}$ in the large $L$ asymptotic expansion (\ref{asympL}).  However,  keeping the summation up to $Q_{-N}\frac{1}{L^N}$, the estimated error rate will be reduced to $\sim\frac{1}{L^{N+1}}$.
So with a suitable choice of $L$ and $N$, we can efficiently calculate  the functional determinant to a desired accuracy  with a relatively small number of $L$.

A generalization of this work to fermi fields should be important.  It can be done by converting the
Dirac operator  into the squared second order form, and then by applying the method developed in the present work. However it should be possible to develop a more direct fermionic approach, studying the coupled first order equations, along the line of the partial wave cutoff method.
Developing a similar partial wave method, to evaluate quantum corrections to the masses of solitons like vortices and magnetic monopoles in gauge theories is certainly an interesting problem.

\acknowledgments
We are grateful to Professor Choonkyu Lee for helpful discussions and for a careful reading of the manuscript.  H. M. thanks the Korea Institute for Advanced Study for hospitality.

\end{document}